\documentclass[prl,twocolumn,showpacs,preprintnumbers]{revtex4}
\usepackage{amsmath,amssymb,latexsym}

\def\be{\begin{equation}}
\def\ee{\end{equation}}
\def\bea{\begin{eqnarray}}
\def\eea{\end{eqnarray}}

\usepackage{graphicx} 

\begin{document}

\title
{Hot-electron relaxation in dense `two-temperature' hydrogen.
}

\author
{M.W.C. Dharma-wardana\email[Email address:\ ]{chandre.dharma-wardana@nrc-cnrc.gc.ca} }
\affiliation
{National Research Council of Canada, Ottawa, Canada, K1A 0R6}

%
\date{\today}

\begin{abstract} 
Recent theories of  hot-electron relaxation in  dense hydrogen or deuterium are  examined
in the light of recent molecular-dynamics simulations as well as various theoretical developments
within the two-temperature model. The theoretical work since 1998 have led to the
 formulation of the $f$-sum version of the Fermi Golden rule formula as the most
convenient  method for the calculation of the rate
of cooling of hot electrons where energy is transferred to cold ions. The attempt to include 
relaxation via the  ion-acoustic modes of the two coupled subsystems, i.e., electrons and ions 
has led to a coupled-mode formulation which has now been established by a variety of formal
methods. However, various simplified calculational models of the system with coupled-modes, 
as well as sophisticated molecular dynamics simulations seem to disagree. It is expected that
coupled-mode calculations which use the simple Coulomb potential $V_{ei}(r)=-|e|Z/r$ for the
 electron-ion interaction within RPA will greatly over-estimate the coupled-mode contribution.
A weak pseudopotential $U_{ei}(r)$ would probably bring the estimated coupled-mode  
contribution to agree with that obtained by simulations. It is also suggested
 that the available `reduced models' have been constructed without much attention to
 the satisfaction of important sum rules, Kramers-Kr\"{o}nig relations etc. We also deal with the
 question of how strongly coupled ion-ion systems can be addressed by an extension of the
 second-order linear response theory which is the basis of current formulations of energy
 relaxation in warm-dense matter systems. These
are   of interest in a variety of fields including hot-electron semi-conductor devices,
inertial-fusion studies of  hot compressed hydrogen, as well as in astrophysical applications.  
\end{abstract}

%
\maketitle
\section{Introduction}
\label{intro} 
Deuterium and tritium mixtures are used in the inertial-confinement approach to
fusion where energy is deposited into an imploding capsule creating a system of
hot electrons and relatively cold ions~\cite{BeneELR17}. The same issues arise
in semiconductor- or solid-state plasmas created using short-pulse
 lasers~\cite{Ng2011}. If a two-temperature
(2$T$)  quasi-equilibrium model can be used, an electron temperature $T_e$ and
an ion temperature $T_i$ are specified where the temperatures are effectively
Lagrange multipliers associated with the conservation of energy in the two
subsystems (electrons and ions) over the relevant time scales; the timescale is
set by the energy-relaxation time $\tau_{ei}(E)$. At sufficiently high
temperatures $T_e$ (i.e, compared to the electron Fermi energy $E_F$), the
energy-relaxation time is proportional to the temperature relaxation time
$\tau_{ei}(T_e)$ and hence the discussion is couched approximately in terms of
temperature relaxation rates $dT_e/dt$, where we assume that the cooler system, (viz.,
the ions) to be attached to a heat bath held at the temperature $T_i$. 
Alternative assumptions can be made, including the use of two heat baths for the
two subsystems, when the physics becomes substantially different.
  
The fusion capsules are made up of an outer ablation layer containing an
admixture of substances that produce high-$Z$ ions, where $Z$ is the mean
ionization (the number of free electrons per ion). For instance, plastic
ablation layers produce $Z \ge 4$ carbon ions
at the compressions and temperatures encountered in the problem. Thus the
simulation of these systems  brings us to the complex question of energy
relaxation of ions of arbitrary charge $Z$ at temperatures $T_i$ interacting
with electrons at an elevated temperature $T_e$. The traditional approach to
this problem, stemming from the days of Landau, is to use a
classical-trajectory approach treating binary collisions among particles, and
allowing for the Coulomb interaction by a `cutoff', leading to the so-called
Coulomb Logarithm (e.g., see Chapter 4,  Ref.~\cite{LandauLif10}). This
approach to energy relaxation (ER) is implemented in the MD-simulations of Hanson and
McDonald~\cite{HanMac83}, while a more extended  theoretical analysis (which
essentially supported the standard results) was given by Boercker and More in
1986~\cite{BoeMoreELR86} and was reviewed in the appendix to Ref.~\cite{DWP-ELR98}.

However, the usual classical trajectory approach is 
rather limited. Since
the particles interact  via potentials whose long-range part is 
Coulomb-like, the essential excitation modes of the system contain not only
particle-particle `binary' interactions, but mostly interactions via their
collective modes, i.e., plasmons. The plasmon modes essentially saturate the
$f$-sum rule (which totals up the number of modes), and hence the particle
 character is subsumed under the collective
modes  which dominate the physics. Furthermore, the collective modes
themselves  interact and produce hybrid modes. In the case of electron-ion
systems, the large mass difference  ($M_i\gg m_e$) implies that electrons
 follow the ions
essentially `instantly', and screen their charge-density fluctuations to create
ion-acoustic modes that  are well known in solids, liquid metals and  in
semi-conductor plasmas. They were also recognized in plasmas already in the 1930s
in the work of Silin {\it et al.}~\cite{Akhiezer75}.
Unlike in electron-hole plasmas or
in semiconductors, the electron-like excitations and ion-like excitations
remain well separated since the ion mass $M_i\gg m_e$. 
 Nevertheless, a complete theory of energy-relaxation in these systems
should take account of  collective modes as well as their coupled modes in a
self-consistent manner. Furthermore, experiments in semiconductor plasmas had
clearly shown that ER-rates of hot-electron cooling were significantly slower
than those predicted by the simple Fermi Golden rule (FGR) based on the
energy transfer from hot-electron  plasmons to cold-ion phonons (ion-density fluctuations).
 Similar, but less
clear evidence existed for slower energy relaxation
in plasmas as well~\cite{NgELR86}. 

As even the FGR calculation
had not been used in ER calculations for dense plasmas, the  present author
attempted to publish such a theory entirely within a quantum approach in 1996,
using the two-particle non-equilibrium Green's functions to formulate
a consistent theory; but
this was rejected by journal referees who held that a two-particle theory to be
inconsistent with the well-established trajectory approach which (they claimed)
clearly implied a `one-particle'  approach for ER in dense plasmas. However,
two years later a longer  paper was published jointly with Perrot~\cite{DWP-ELR98}
 presenting the FGR as well as the coupled-mode expressions for ER. 
The present-day reader of
 that the paper  may note a (seemingly irrelevant) running discussion about the
inapplicability of the one-body propagator to energy relaxation problems, due
to the earlier abortive debate with journal referees well anchored in classical
trajectory calculations. Today most workers accept that the
excitation modes of the two-particle propagator and their damping hold the key
to ER rates.

The analysis of Ref.~\cite{DWP-ELR98} suggested that simple estimates based on
binary-collisions with Coulomb cutoffs, or more systematic treatments using the
Fermi golden rule, but neglecting the screening of the ion excitations by the
electron excitations (i.e., coupled modes, {\it cm}) would predict ER-rates larger than
what is physically correct. In effect, the `neutral-pseudo atoms' (NPA)  formed
by the ions screened by the hot electrons were objects intermediate in
temperature  to $T_i, T_e$ and hence the ER-rate is significantly slowed down.
The hybrid plasmons made up of electron-density excitations as well as ion-density
excitations are the coupled modes. 
The effective coupled-mode temperatures $T_{cm}(\omega)$ are also dependent on the mode
energy $\omega$; thus for instance, when $\omega\to \omega_e$, where
$\omega_e$ is the electron plasma frequency,  then $T_{cm}\to T_e$. However,
unlike with room temperature materials, obtaining accurate ER-rates for
warm-dense matter (WDM) systems is even more difficult than obtaining accurate
electrical conductivity data for WDMs. Calculating reliable conductivities themselves
for WDM, e.g., via density-functional theory (DFT), molecular dynamics (MD)  and the
Kubo-Greenwood formula requires, even for a simple metal like sodium, a simulation
involving over a 1000 atoms and many $k$-points, of the order of
 50-60 points~\cite{Pozzo11}.
Hence the challenge for carrying out reliable simulations of energy relaxation is
even greater.

Meanwhile, plasma kinetic-theory methods that are more familiar to the WDM
community were deployed to study the ER-rates in  2$T$ quasi-equilibrium
systems. Unlike the Lenard-Balescu method, the Klimontovich approach
pays attention to both kinetic and potential energy terms in the
dynamics. Rosenfeld had derived the coupled-mode equations of ER using
Klimontovich's methods~\cite{pcom} 
and found its numerical implementation quite demanding. At the time, the present author was 
looking into implementation of the $f$-sum rule and other sum rules in ER-calculations.
Hence  we began to
examine them as a means of simplifying the FGR and the {\it cm}-formulations. Not surprisingly, 
it was not possible to apply the $f$-sum rule to the coupled-mode problem but
 the FGR calculations
were greatly simplified~\cite{HazakELR01}. Some time later the Lenard-Balescu (LB) equations
were used by  Gericke {\it et al.}~\cite{GerELR05,VorGerSchELR10} who also arrived
at coupled-mode effects within an LB formulation. In such kinetic theories,
and in Zubarev's real-time green's function method, since the intermediate frequency
integrations are `already done', vertex corrections and self-energies are
partitioned in a a different way  and it is not easy to
bring in our experience from standard many-body theory to the kinetic-theory
methods. On the other hand, since we are dealing with nonequilibrium
systems where many things are
 unclear,
 the reexamination of the problem via a variety of methods is necessary.

  Nevertheless, at the
level of coupled modes (but without vertex corrections, LFCs etc.), 
the various theories
are in agreement. The disagreement seems to reside in the question of 
when {\it cm} become relevant and possible differences in various 
different theories themselves, e.g., as
discussed in
Daligault {\it et al.}~\cite{DaligDia09}. Usually, in a many-body theory, whether {\it cm}
is important or not is
taken care of by the theory itself thorough the interplay between the real and
imaginary parts of the excitation modes. But it is relevant in computations
for avoiding the  heavy cost of implementing the full theory. 
Given the difficulty of obtaining experimental ER-rates, theorists have turned
to non-equilibrium molecular dynamics (MD) simulations to determine ER-rates
for comparing theory with numerical
`experiments'~\cite{BeneELR17,DaligDia09,MuriDW-elr08,Glosli-MMLL08,HanMac83}.
This is non-trivial since the usual Born-Oppenheimer approximation
is no longer available to simplify the simulations.
The simulation results regarding coupled modes seem to lead to
 contradictory conclusions. Furthermore, some kinetic-theory  analyses
have given ER-rates larger than from FGR, while most theorists agree with our
earlier analysis~\cite{DWP-ELR98} that the presence of coupled modes reduces the
rate of energy relaxation, using quite improved numerical computations. 

The objective of this review is to
examine these conclusions and shed some light on the implementation of coupled
modes in ER rate calculations. The work of Daligault {\it et al.} is interesting
since the simulation of ER in a plasma of like charges (`repulsive hydrogen') removes the
many uncertainties associated with using the so-called quantum statistical potentials
(QSP), e.g., as used by Hansen and McDonald in Ref.~\cite{HanMac83},
 to control attractive electron-ion interactions.
It will be shown that the equations used by Daligault and Dimonte (DD) are in 
agreement with the equations given by the author and Perrot in Ref.~\cite{DWP-ELR98},
 although DD have claimed it to be otherwise. This agreement holds both in the quantum
 regime, and in the classical regime. The contrary conclusion of DD is partly due 
 to shortcomings in our proof reading,
 although the information is
clear from a number of explanatory subsections
and an appendix.
Furthermore, their different point of view regarding the description of
 two-temperature ultra-fast
matter has led to their  `self-consistent' formulation
which seems to us to be inappropriate unless a
 two-thermostat 2$T$ system is envisaged, as will be discussed below.

\section{Energy relaxation by plasmon modes.}
\label{ER.sec}
A Coulomb system with free electrons and ions can have electron plasmons and ion plasmons, and
 their coupled modes. The ion-electron coupled mode manifests as the ion-acoustic modes
of the plasma.
 Some authors have used Gordeev's criterion~\cite{Gord54} for the existence of
well-defined propagating ion-acoustic waves ($T_e\gg T_i$) for determining the
relevance of coupled modes (ion-acoustic waves)  to ER.  Vorberger
 {\it et al.}~\cite{VorbGeriELR09} have suggested more specific limits ($T_i\le 0.27
ZT_e$) involving the mean ionization $Z$ of the ions. The mean ionization is
simply the number of free electrons per ion, and is a well-defined physical
quantity although some authors have (incorrectly) questioned
 the very concept of a mean ionization~\cite{PironBlenski11}.  The
criteria of Gordeev, or Vorberger {\it et al.} attempts to minimize the
 damping of the ion-acoustic mode,
i.e, to have minimal overlap between the phase velocities of the ion-acoustic
waves and the particles, ensuring that a minimally damped ion-acoustic wave
propagates in the plasma. However,  in our view these are  precisely the
conditions where the ion-acoustic waves do not participate significantly in energy
relaxation.  Hence  MD-simulations which focus
on this regime should show negligible
contributions to the ER rate from coupled modes.  In fact, when ion-acoustic
modes participate in ER, then they are likely to be damped by the very
relaxation process which is due fundamentally to the electron-ion interaction
which also causes the coupled mode.

The Gordeev criterion $T_e\gg T_i$, with $\Delta T=T_e-T_i$ large attempts
to reduce the damping of the ion-acoustic modes, but its
action is somewhat like throwing the baby out with the bath water.
 While the damping in the ion-acoustic mode decreases
as $T_e$ is increased, the ion contribution to the coupled mode also decreases
rapidly, and the coupled mode simply becomes a product of two independent modes
at higher $T_e$. Another factor that has to be considered is the energy
relaxation time. The relaxation time is proportional to $\Delta T$, and hence,
for sufficiently large $T_e$, the relaxation time may be too short for a 
collective mode to be formed, unless a system with two thermostats is envisaged.
 That is, the very high $T_e$ limit reduces to the Landau
Spitzer limit, as evident from Hazak {\it et al.}~\cite{HazakELR01}. The relevant
regime for ER via coupled modes is the overlap region of the
very low-frequency regime of electron excitations and
the high-frequency range of ion excitations, together with the need for sufficiently
long ER times since coupled modes need a certain amount of time to build up
and dissipate.  The latter  problem 
does not arise if the system being studied is controlled by two thermostats, where
the upper thermostat maintains the electrons at the steady state $T_e$, while the
lower thermostat maintains the ions at $T_i$. Such a system can be realized in practice
if hot electrons are pulse pumped into the conduction band of a semiconductor,
while the ion lattice is coupled to a thermostat. In WDM systems, short-pulsed lasers
pulsed at an appropriate rate can raise the electrons to $T_e$, while the ions
remain at their initial state for short time scales $t<\tau_{ei}$. Such 2$T$
 ultra-fast matter (FM) can be studied optically with probe lasers deployed
  immediately after the pump pulse, with a delay exceeding the subsystem equilibration
times $\tau_e,\tau_i $ which serve to set up the subsystem temperatures $T_e,T_i$. 
Only one thermostat (for the ions) is assumed in such studies.

In the following we examine the FGR and coupled mode formulae, keeping in mind the
LB-kinetic theory results as well as the MD-simulations results for ER.\\

\subsection{A review of ER-models at the Fermi Golden rule level.}
\label{review-fgr.sec}
In order to compare and contrast the available theoretical models,
we summarize the basic theory for our convenience.

Assuming that $T_e > T_i$ to be specific, ER occurs via energy transfer from
the excited modes of the electron sub-system to the cold modes in the ion
subsystem which may be assume to be connected to a heat bath without loss
of generality. We do not assume a heat bath at the upper temperature, but
assume that the experiments are done duing timescales $(\tau_e,\tau_i)\ll t \ll \tau_{ei}$.

 The spectral densities of the modes of the species $j=e,i$ are given by the
spectral functions $A_j(q,\omega,T_j)$. The spectral functions are
essentially the dynamic structure factors $S_j(k,\omega)$ of
each subsystem, accessible experimentally using optical experiments.
We use only one subscript, e.g., $A_j$ when we mean $A_{jj}$, when there is
no ambiguity.
The spectral functions  are given by
the imaginary parts of the corresponding dynamic response functions
$\chi_{j}(\vec{k},\omega)$, e.g., Eq.~(16) of Ref.~\cite{HazakELR01}.
 If the
electron-ion interaction $V_{ei}(r)$ is {\it weak}, then the Fermi Golden rule and
other methods based on linear-response theory can be used. For this purpose the
electron-ion interaction $V_{ei}(q)$ cannot be taken as $-ZV_q, V_q=4\pi/q^2$
except in the Gell-Mann$-$Brueckner (quantum) limit. Coulomb collisions, be they classical
or quantum, require regularization both at short range, and at long range.
 Hence the use of a pseudopotential $U_{ei}(r)$ which behaves as the
Coulomb potential for $r>r_c$, and regularized for $r<r_c$, where $r_c$
is effectively the core radius of the ion is needed. Even with hydrogen, although
it has no bound-electron core for the conditions of interest, such a form
is needed since the electron pile up very close to the nucleus is highly
nonlinear. 
The long-range of the potential will be corrected automatically in the theory by
screening effects of other particles. The construction
of these pseudopotentials using the NPA-average atom (AA) model 
was discussed in sec.II of Ref.~\cite{DWP-ELR98}, and will be summarized here
for the convenience of the reader (see sec.~\ref{NPA.sec}). In kinetic-equation methods
a short-range local-field-correction (LFC) $G_{ei}$
is invoked via the factor  $\{1-G_{ei}(k,0)\}$ estimated from HNC equations, 
to provide some sort of pseudopotential. 

\subsection{\bf Fermi Golden rule results.}
\label{FGR.sec}
The ER rate evaluated within the Fermi golden rule, $R_{fgr}$
 can be expressed in terms of
the response functions of the plasma as given in Eqs. (4)-(7)
 of DWP, Ref.\cite{elrDW01}, and Eq.~(12) of Hazak {\it et al.}, Ref.~\cite{HazakELR01}.
The imaginary part of the response function gives the mode spectrum, or
spectral function $A_j(k,\omega)$.
\bea
\label{spectral.eq}
A_j(q,\omega,T_j)&=&-2\Im \chi_{jj}(q,\omega,T_j)\\
\label{chi.eq}
\chi_{jj}(q,\omega,T_j)&=&\chi_{jj}^0/\{1-V_{jj}(q)[1-G_{jj}]\chi_{jj}^0\}
\eea
Furthermore,
\bea
\label{cothform}
R_{fgr}&=&\frac{\delta E}{\delta t}=2\int \frac{d^{3}k}{\left( 2\pi \right) ^{3}}
\frac{\omega d\omega}
{2\pi}(\Delta N) F_{ei}\\
\Delta N&=&N(\omega/T_e)-N(\omega/T_i)\\
N(\omega/T_j)&=&1/[\exp^{\omega/T_j}-1] \\
\Delta N&=&(1/2)\{\coth(\omega/2T_e)-\coth(\omega/2T_i)\}\\
F_{ei}&= &|( U_{ei}(k)| ^{2}\Im\left[ \chi_e(\vec{k},\omega )\right] 
\Im\left[ \chi _{i}\left( \vec{k},\omega \right) \right]
\eea
The plasmons are bosons, and hence Bose factors $N_j(\omega/T_j)$ occur
in $\Delta N$, which is the excess-plasmon population  which drives the energy flow.
In the above $\delta E/\delta t$ is the rate of change of the energy of the
system, for time steps $\delta t$ significantly greater than the
equilibration times $\tau_e, \tau_i$ which establish $T_e$ and $T_i$ of each
subsystem. The relaxation of the whole system is determined by $\tau_{ei}$ such
that $\tau_{ei}>>\tau_i>\tau_e$. For brevity we write $\delta E/\delta t$  as
$dE/dt$. The spherical symmetry of the plasma (i.e., not for solid-state plasmas)
is used to write scalars $q,k$
instead of $\vec{q},\, \vec{k} $ to simplify the notation. The  non-interacting
(one-component) response function $\chi^0(q,\omega,T)$ at arbitrary degeneracies
 was given by
Khanna and Glyde\cite{KhGly78}, and are used here. This reduces to the Lindhard
form at low-$T/E_F$ and the Dawson form at high $T/E_F$. The full response
function $\chi_j(q,\omega,T)$ with $j = e,i$ uses a $T_j$-dependent local field
corrections, e.g.,   $G_{ee}(k)$~\cite{PDWXC} derived from the finite-$T$
electron-electron exchange-correlation (XC) functional of DFT. The full
$k$-dependence is given in Ref.~\cite{PDWXC}. The $\omega$-dependence
 of the LFC is needed only in the evaluation of coupled modes (see~\ref{sec.lfc}).
 If $T_e,T_i$ are {\em both} sufficiently large so that $\Delta N\to (T_e-T_i)/\omega$, and if
the electron chemical potential $\mu_e \le 0$, useful analytical approximations
become available. The possibility of unequivocally extracting a
temperature-relaxation time $\tau_{ei}$ from the relaxation rate exists only in this
regime, as was well-kown in ER studies in solid-state semiconductors.  Neglecting 
interactions,
 $E$ becomes the kinetic energy. Using
non-interacting classical forms for $\Im\chi^0_j(k,\omega)$ in
Eq.~\ref{cothform} we obtain the well known Landau-Spitzer (L-S) form for the
temperature relaxation time $\tau$ or $\tau_{ei}$.
 For the L-S form we set $U_{ei}$ to be the Coulomb
interaction $V_{ei}(k)$. Then,
\bea
1/\tau&=&\frac{2}{3n}\omega_{p_e}^2\omega_{p_i}^2\left[(2\pi T_{ei})/m_{ei}\right]^{-3/2}{\cal L}\\
 {\cal L} &=& \log(k_{max}/k_{min}) \\
  T_{ei}/m_{ei}&=&T_e/m_e+T_i/M_i, \;\;\; \omega_{p_j}^2=4\pi n/m_j
\label{coullog}
\eea
Here $\omega_{p_j}$ is the plasma frequency of the  species $j =e,i$. The
effective temperature and the effective mass of the colliding pair are $T_{ei}$
and $m_{ei}$, with $T_j$ in energy units. ${\cal L}$ is the ``Coulomb
logarithm''. It depends on $k_{min}$ and $k_{max}$, i.e., momentum cutoffs (or
impact parameters) used for modeling the unscreened Coulomb collision. If
interacting response functions (e.g., RPA and beyond) are used, single-particle
modes become replaced almost completely by plasmon modes, and the interactions
become dynamically screened. However,  ``static screening'' emerges when the
$f$-sum rule is used to reduce the frequency integrations using the fact that
$m_e/M_i$ is very small and hence electrons follow the ions  `instantly'.

Hence a well-controlled procedure to do the $\omega$-integration is to
 exploit the $f$-sum
rule\cite{HazakELR01}. Then ion dynamics are automatically preserved. Writing
$\Delta=(T_e-T_i)$, Eq.~\ref{cothform} simplifies to:
\bea
\frac{1}{\Delta}\frac{d\Delta}{dt}&=&\frac{2}{3n}\omega _{p_i}^{2}\int_0^\infty\frac{2}{\pi}
\left[ \frac{\partial }{\partial \omega }\Im\chi ^{ee}\left(
k,\omega \right) \right] _{\omega =0}dk
\label{fgrBA}
\eea

Hence only the static form of the electron response function is needed, and the calculation
 is reduced to a simple $k$-integration.

\subsection{Nearly analytic form for use with systems where the electron chemical
potential is less than zero.}
If the electron chemical potential $\mu\sim0$, or dips below zero, the degeneracy effects
of the plasma can be adequately treated by retaining terms in $\chi_e(k)$ only up to $\hbar^2$.
In most of the regime of interest to the simulations of Ref.~\cite{BeneELR17} and
some of the simulations reported in Ref.~\cite{DaligDia09}, one can
approximate  $\Im\partial\chi^{ee}/\partial\omega\,|_{\omega =0}$  as:
\bea
\Im \partial\chi_{ee}/\partial\omega\;|_{\omega=0}&=&\frac{\Im\partial\chi^0_{ee}/
\partial\omega\;|_{\omega=0}}
{\{1+k^2_{sc}/k^2\}^2}
\label{derivchi}
\eea
The $k\to 0$-local field
correction, $G_0^{ee}$ at arbitrary degeneracy\cite{PDWXC} can also be included
in $k_{sc}$ {\it via} the following definitions.
\bea
(k^0_{sc})^2&=&\frac{2}{\pi}(2T)^{1/2}I_{-1/2}(\mu^0_e/T_e)\\
I_{\nu}(x)&=&\int_0^{\infty}\frac{dyy^\nu}{e^{y-x}+1},\, \nu\ge-\frac{1}{2} \\
k_{sc}&=&k^0_{sc}\left[1-G^0_{ee}\right]^{1/2}
\label{screenkc}
\eea
In approximating $\Im\partial\chi^{ee}/\partial\omega$ we retain quantum
corrections to second order in $\hbar$, as displayed explicitly below, correcting
a typographical error in Ref.~\cite{CDW-elr08}
\be
\Im\chi^0 _{ee}=-(\frac{\pi}{2T_e})^{3/2}\,\frac{2n\omega}{\pi k}
e^{-\frac{1}{2T_e}\{\frac{\omega^2}{k^2}+\frac{\hbar^2k^2}{4}\}}
\frac{\sinh(\hbar\omega/2T_e)}{\hbar\omega/2T_e}
\ee
Then Eq.~\ref{fgrBA} can be reduced to the form:
\bea
\label{newresult1}
1/\tau&=&-\frac{2}{3n}\omega_{p_e}^2\omega_{p_i}^2\{2\pi(T_{ei}/m_{ei})\}^{-3/2}{\cal Q}\\
{\cal Q}&=&\frac{1}{2}\left[e^pEi(p_e)(p_e+1)-1\right]\\
p_e&=&k_{sc}^2/(8T_e)\\
Ei(x)&=&\int^{\infty}_x exp(-t)dt/t
\eea 
The exponential integral~\cite{gradstein}, $Ei(x)$ of
Eq.~\ref{newresult1}, is evaluated numerically via standard subroutines. Thus
we see that the ``Coulomb factor'' ${\cal Q}$ is exactly analogous to the
``Coulomb logarithm'' of Eq.~\ref{coullog}, but without {\it ad hoc} cutoffs.
${\cal Q}$ contains leading-order quantum corrections, ion-dynamics and
electron screening. The expression for ${\cal Q}$ should be compared with a
similar expression given by Brown {\it et al.}~\cite{br-sin} which gives nearly equivalent
numerical results, and hence reveal the physics content of the Brown {\it et al.}
result. At high $T_e$, this result
approaches the L-S form more rapidly than ${\cal Q}$~\cite{CDW-elr08}.

It is likely that Eq.~\ref{newresult1} should be adequate for evaluating most of 
the cases
of H-plasmas studied in Ref.~\cite{BeneELR17}.

\section{Energy relaxation via coupled modes}
\label{cm.sec}
The interactions between the ion modes  and electron modes lead to ion-acoustic
modes (coupled modes). It is seen in Fig.1(a) of Ref.~\cite{DWP-ELR98}  that electron
density fluctuations in the electron density (represented by a shaded loop and denoted
by $\chi_{ee}$ ) modify the bare Coulomb interaction $V_{ee}(q)$ to give a screened 
interactions obtained by summing a geometric series of such polarization loops. 
This resummation does not take account of other classes of diagrams (e.g., ladder sums)
which may dominate under other conditions of density and temperature.
Similar processes arise from the ion density fluctuations $\chi_{ii}$, modifying
the bare ion-ion interaction.  Furthermore, processes involving
 both types of loops occurring arbitrarily become possible,
as seen from  Fig.2 (c) and Fig.3 (d) of Ref.~\cite{DWP-ELR98}. It
is also evident that modifying the ion-ion interaction
 line $V_{ii}$ by `screening it' with ions by writing it as
$V_{ii}/\epsilon_{ee}(k,\omega)$ produces no effect as those terms
are already included, although some authors have suggested such
nonpermitted `extensions' using quantum-kinetic equation methods.

These particle-hole processes modify all interactions 
including the electron-ion interaction $U_{ei}(q)$ which determines the
relaxation from hot electrons to cold ions. The  charge excitations couple together,
 just as  two harmonic oscillators couple together to give combination  modes.
If the system were in equilibrium ($T_e=T_i\to 0$) the standard $T=0$ Feynman rules can be
used to evaluate these diagrams trivially. It turns out that the result obtained from a more
sophisticated evaluation (e.g., using non-equilibrium Martin-Schwinger-Keldysh theory)
has the same algebraic structure  as
that obtained from a simple analysis (formally similar results can also be
obtained using  various kinetic-equation methods).
However, the result in a given order in perturbation theory is usually expressed in terms 
of lower-order quantities. At that point, one may replace the lower order quantities
(e.g., propagators, spectral functions, denominators, LFCs, etc.) by fully renormalized
`self-consistent' quantities by further resummations or insertions of vertex
corrections, self-energies etc. However, this involves pitfalls in the context of non-equilibrium
systems, not only with diagrammatic methods, but even more so with
kinetic-equation methods (as discussed below). Even at the level of single-particle band-structure
calculations at $T=0$, one is reminded of GW calculations. If they
are made `more self-consistent' by adding
vertex corrections, self-energy insertions  etc., they give worse
 results because such seemingly `more self-consistent'
improvements may  not actually give a conserving approximation.

The energy relaxation rate $dE_e/dt$ via {\it cm} is given by
Dharma-wardana and Perrot (DWP) in Eq. [50], Ref.~\cite{DWP-ELR98}. 
However, the shortcomings in our notation and in our proof-reading
seem to have confused
a number of readers. The electron-ion interaction in the numerator is 
correctly given as $U_{ei}$ and discussed in detail in various parts of the paper;
it is not identical to the Coulomb interaction $V_{ei}$ but reduces to
a Coulomb potential $V_{ei}$ only for large $r$ (or small $k$). Nevertheless many writers
have simply replaced our  $U_{ei}$ by $V_{ei}$ in their restatement of
our work, e.g. in  Eq. (25) of Daligault and Dimonte (DD), Ref.~\cite{DaligDia09});
 they then concluded that the
theory fails in the classical regime. The spectral functions used in Eqs.
 (47)-(50) of DWP have also perhaps been a source of confusion although they are
clearly defined and the typographical errors etc., sort themselves out if
one re-derives Eq.~(50) from Eq.~(47) of DWP.  We give below  Eq.(50) of DWP
for energy relaxation
via coupled modes, with the arguments $\vec{k},\omega$ 
 suppressed for brevity.
\begin{eqnarray}
\label{cm-dwp.eq}
dE_e/dT&=&\int \frac{dk^3}{(2\pi)}^3 \frac{\omega d\omega}{2\pi}|U_{ei}(k)|^2 \mathcal{R}\\
\mathcal{R}&=&-(1/2)\frac{A_eA_i\Delta N_{ei}}{|1-U_{ei}^2(k)\chi_e\chi_i|^2}\\
\label{spec.eq}
A_j(k,\omega)&=&-2\Im \chi_{jj},\; \mathcal{F}_{jj'}=1-G_{jj'}\\
 \chi_{jj}&=&\chi_{jj}^0/D_{jj} \\
 D_{jj}&=&[1-V_{jj}\mathcal{F}_{jj}\chi^0_{jj}].
\end{eqnarray}
The Coupled denominator emerges explicitly if one brings down the denominators
for $\chi_{jj}$ and incorporate them into the factor
 $ \{|1-U_{ei}^2(k)\chi_e\chi_i|^2\}$.
\begin{eqnarray}
\label{cmD.eq}
\mathcal{D}&=&D_{ee}D_{ii}-D_{ei},\\
\label{ocD.eq}
D_{ei}(k)&=& |U_{ei}|^2(k)\chi_{ee}^0\chi_{ii}^0\\
|U_{ei}(k)|^2&\simeq&(1-G_{ie})(1-G_{ei})V_{ei}(k)^2
\end{eqnarray}
The last equation is only approimate, since we do not determine $U_{ei}$ from
the LFCs.
The $A_e, A_i$ used in the above equations
are the spectral functions based on independent subsystems, as defined in Eq.~\ref{spectral.eq}
in terms of $\chi_{jj}$ which has a simple denominator. If the spectral functions
are defined in terms of the two-fluid model using a coupled-mode denominator $\mathcal{D}$
then the spectral function is denoted by $A_{cm}^j$ in Eq. (47) of DWP, together with
a $\Delta N_{cm}$ for the excess Boson population that drives the energy relaxation. However
Eq. (47) is transformed to eq. (50) of DWP, which is Eq.~\ref{cm-dwp.eq} given above. This
contains only the independent-subsystem spectral functions $A_j,\; j=e,i$.  
  
We give below the  coupled mode form 
given by Daligault and Dimonte~\cite{DaligDia09}. The quantities used in the formulation
by DD are marked with an asterisk, $^*$, to distinguish them from our definitions.
\begin{eqnarray}
\label{cm-DD.eq}
dE_e/dT&=&\int \frac{dk^3}{(2\pi)}^3 \frac{\omega d\omega}{2\pi}|V_{ei}(k)|^2 
[1-G^*_{ei}]\mathcal{R}^*\\
\mathcal{R}^* &=& -(1/2)\frac{A^*_eA^*_i\Delta N_{ei}}{|1-U^*_{ei}(k)^2\chi^*_e\chi^*_i|^2}\\
\label{chi-DD.eq}
|U_{ei}^*(k)|^2&\simeq&(1-G^*_{ie})(1-G^*_{ei})V_{ei}(k)^2\\
A^*_j&=&-2\Im \chi_{jj}^*\\
 \chi_{jj}^*&=&\chi^0_{jj}/[1-V_{jj}(1-G^*_{jj})\chi^0_{jj}]
\end{eqnarray}
The equation given by Daligault {\it et al.} has a numerator
$|V_{ei}(k)|^2\{1-G^*_{ei}\}$ which is in second order only in the Coulomb part of the
potential, while the denominator contains 
$|V_{ei}(k)|^2\{1-G^*_{ei}\}\{1-G^*_{ie}\}$.

It is actually necessary to use $\omega$ dependent LFCs in these equations. However,
most of the work so far has replaced $G_{jj'}(k,\omega)$ by $G_{jj'}(k,0)$. 
We examine these equations in more detail below to clarify the differences between DWP
and DD formulations.

\subsection{Interaction potentials, local-field corrections and the $2T$-equation of state.}
\label{sec.lfc}
The DD-equations use Coulomb potentials $V_{jj'}$ corrected by their LFCs $(1-G_{jj'}^*)$
(taken in the static approximation). It appears that they are calculated from the
 Ornstein-Zernike equation for
a two-temperature plasma at $T_e, T_i$ `self-consistently'. Similarly, it
may be  that even $\chi_{ii}^*$ is similarly  `self-consistent'. This self-consistency
is {\it deliberately not included} in the DWP formulation of $2T$-quasiequilibrium systems. 
As explained in Ref.~\cite{DWP-ELR98}, and discussed at length in the Appendix
there,  in regard to the quasi-equation of state. We consider a system of electrons and ions
 both initially at equilibrium  at $T_i$, when the electrons are very rapidly raised to a
temperature $T_e$ by a short-pulse pump laser. The objective is to describe the system within
time scales shorter than $\tau_{ei}$ such that {\it the ions have had no time} to relax to an
equilibrium state. The ion-ion LFCs etc., remain `frozen' at their initial values
 $G_{ii}(T_i=T_e)$,  $g_{ii}(r,T_i=Te)$, $S_{ii}(k,T_i=T_e) $ etc.
Hence they are {\it not what is self-consistent} with
the new-electron distribution at $T_e$. 

Thus, in our view, the use of a `self-consistent'
$G^*_{ii}$ etc., is not consistent with the assumptions of the quasi-equilibrium 2$T$ state
normally generated in laser experiments. 
On the other hand, the electrons  readjust to the new conditions in femto-second time scales; 
the $G_{ee}$ become $G_{ee}(T_e)$ in the external field of the ions still specified
by the {\it initial}  $g_{ii}(k,T_i=T_e)$ etc. Thus our LFCs $G_{ee}$ are also different
from the $G_{ee}^*$ used by
DD. The $G_{ee}$ used in the DWP calculation was constructed in linear response to the
unmodified $g_{ii}(T_i,T_e=T_i)$ initial state of the ion distribution which was generated from
the electron-ion pseudopotential $U_{ei}(k)$ calculated at $T_e=T_i$, i.e., the initial state when
the energy was deposited by an ultra-fast laser pulse. The initial state $U_{ei}(k), g_{ii}(r)$
 etc. were calculated using the NPA+MHNC procedure (sec.~\ref{NPA.sec}).

The validity of the NPA+MHNC procedure used
by us has been checked over the years, and also in recent calculations
 of the 2$T$-EOS for Al, Na and Li (at
normal density and under some compressions). There
the Helmholtz  2$T$-Free energy $F(T_e,T_i)$ and derived quantities like the
 internal energy $E(T_e,T_i)$, pressure $P(T_e,T_i)$  were evaluated using the NPA+MHNC as
 well as DFT+MD, and shown to agree very well in the range where DFT+MD
 could be implement~\cite{Harb17Phon} (the comparison is limted to low $T/E_F$ since
DFT+MD using codes like VASP or
 ABINIT can only be implemented up to about $T/E_F<0.5$  when the 
number of electronic states that have to be included becomes prohibitive). Furthermore,
 the validity of such codes in such regimes had not been addressed up to then. Our
 NPA+MHNC calculations mutually validate the procedures used, and the NPA+MHNC could be
 seamlessly used for arbitrarily higher $T$.  In the above procedure, the  $2T$ static
 electron quantities like $g_{ee}(r), S_{ee}(k),\chi_{ee}(k), g_{ei}(k)$ are readily
 available from NPA as well as classical-map HNC calculations~\cite{PDWXC}.  

In contrast, $g_{ii}(r)$ and other quantities used in DD, and possibly in Benedict
 {\it et al.}~\cite{BeneELR17} may be quantities derived from a self-consistent
 two-component 2$T$-simulation where two thermostats are assumed to maintain the
systems in equilibrium at $T_i$ and $T_e$. The physical quantities that enter
 into the two-thermostat problem are different from those of the one thermostat problem.
Clearly, the DWP equations and the DD equations closely agree when applied to such
 systems with
 two thermostats, or when applied to systems with the ions clamped at the initial state,
 when appropriately
 computed after recognizing what system is being studied.

It should be noted that the direct-correlation functions $c(k)$ of Onstein-Zernike (OZ) theory are
related to the LFCs used with the neutral-pseudo-atom potentials, as elucidated by
Perrot, Furutani and Dharma-wardana~\cite{Furutani90}. Using
 Eq.(24]) of Ref.~\cite{Furutani90} or other equations, it is seen
that:
\begin{equation}
\label{lfc-cr.eq}
\mathcal{F}_{jj'}=1-G_{jj'}(k)=1-\tilde{c}_{jj'}(T/V_{jj'})
\end{equation}
where $\tilde{c}_{jj'}$ is the short-ranged direct correlation function of OZ theory.

We have used a different notation $U_{ei}$ distinguishing it from the Coulomb potential, and discussed its calculation from
 a full quantum
Kohn-Sham equation (and fitted to an extended Heine-Abarankov pseudopotential for convenience),
 as given in Eq. (60) of DWP. But our $U_{ei}(k)$ is set to a bare Coulomb potential by DD
 perhaps to be in line with other Coulomb interactions. Then
DD claim(Ref.~\cite{DaligDia09}) in item (b) just before their conclusion that
``Unlike our model, the
DWP model diverges logarithmically at large k ... the integrand scales like $1/k$ at large
 $k$....''.
 The DWP model at the FGR level,
and at the {\it cm} level are free of such large-$k$ (or small-$k$) divergencies, and includes both
 quantum and classical short-range (large-$k$) corrections in $U_{ei}(k)$ appropriately,
 satisfying the Friedel sum rule, and even dealing correctly with bound-state formation in the quantum
case~\cite{CDW-elr08}.
 
A more detailed look at $U_{ei}(k)$, calculated via the Kohn-Sham equations of the NPA
model~\cite{Pe-Be, eos95} will be presented below. Although $U_{ei}(k)$ is an `all-order'
interaction, it has been derived to be compatible with linear  response theory where
interactions are treated to second-order in the screened interactions, and hence we believe
that the inclusion of $U_{ei}$ in the numerator is consistent as long as
higher-order terms are not included. Our experience with electrical conductivity calculations
using $U_{ei}(k)$ in the Ziman formula confirm this conclusion.

\subsection{\bf The electron-ion pseudopotential derived from the Neutral-Pseudo-atom model.}
\label{NPA.sec}
The electron-ion interaction is given as a weak pseudopotential having the form
$U_{ie}(q)=-ZV_qM_q$, where $V_q$ is the Coulomb potential $4\pi/q^2$,  and
$M_q $ is the `matrix element' or form factor that regularizes the interaction to 
be compatible with second-order perturbation theory. The appropriate pseudopotentials
are derived from density-functional theory (DFT) calculations using the NPA-average atom
model as given by Perrot and Dharma-wardana (PDW)~\cite{Pe-Be, eos95}. The 
PDW model is different from a  number of other available AA models, e.g.,
Purgatorio, MUZE~\cite{Murillo13} etc., which confine the electrons to
 a Wigner-Seitz sphere, and
lead to several definitions of the mean ionization $Z$ which disagree with each other
especially at high densities and low $T$.  Blenski {\it et al.} find that the
estimate of $Z$ in their model also leads to difficulties at low $T$ and normal densities,
as they illustrate via the case of aluminum~\cite{Blenski13}. In our codes the
free electrons are not confined to the Wigner-Seitz sphere, and the model is
valid at low- or high $T$, and at all densities except when clustering effect
become important. However, at the regimes of $T$ and
ion density $\bar{\rho}$
considered by Ref.~\cite{BeneELR17}, all AA models for the calculation of the
free-electron charge-density at a nucleus, viz., $\Delta n_f(r)-\bar{n}$  and
 the associated free-electron density $Z$ per ion should be quite reliable.

The integral of $\Delta n_f(r)$ calculated from the Kohn-Sham equation
 extending over the
the whole of space  (i.e. up to $R_c=10 r_{ws}$ , in our codes) yields $Z$
(the ionic charge) without ambiguity. The exact procedure for the
determination of the mean ionic charge $Z$ to satisfy the Friedel sum rule etc.
is discussed in more detail in Ref.~\cite{cdw-carbon16}.
 The $e-i$ and $i-i$ interaction potentials are given by
\begin{eqnarray}
\label{pseudo-ppot.eq}
U_{ei}(q)&=& \Delta n_f(q)/\chi_{ee}(q,0)\\
U_{ii}(q)&=&Z^2V_q + |U_{ei}(q)|2\chi_{ee}(q,0).
\end{eqnarray}
Hence the static electron response function $\chi(q)$ and the Kohn-Sham density
pile up around the ion completely define a weak local ($s$-wave) pseudopotential
and the ion-ion pair-potential, with no {\it ad hoc} parameters. This
linear-response pseudopotential $U_{ei}(q)$ can be legitimately used in the
ER-calculations using linear response functions etc., as needed in the FGR and
coupled-mode calculations. There we use $U_{ei}(q)$ to denote the electron-ion
pseudopotential. Thus  short-range corrections of the form
$(1-G_{ei}^*)$ introduced in the LB-type kinetic-equations at an
unknown level of consistency (or `self-consistency')
are also contained in the DWP ER-rate calculation in a form adapted for linear response,
although containing non-linear DFT corrections.
\subsection{The `meaning' of the two-temperatures in two coupled  subsystems.}
\label{temperature.sub}
The Hamiltonian of the system is made up of $H=H_e+H_i+H_{ei}$. Temperature is a quantity which
is not represented by an operator in a simple Hilbert space, but has a meaning only
in quantum statistical mechanics where the system is attached to a heat bath. The temperatures
 $T_e,T_i$ are the Lagrange multipliers associated with the conservation of $H_e, H_i$ relevant to
the time scales of the study $t<\tau_{ei}$. If only the ions are thermostated, then $H_e$
is conserved for timescales $t< \tau_{ei}$.  Other statistical quantities like $Z, \mu_e, \mu_i$,
are also Lagrange multipliers for the conservation of global charge neutrality and the conservation
of particle numbers. However, in writing down the partition function, or in implementing the
HNC equations for such a $2T$-two-component system, the question of  what temperature to use for the
$H_{ei}$ term can arise. This question is meaningless in ultra-fast matter where $H_{ei}$
drives the time evolution in a dissipative manner. The 2$T$-implementations of the NPA+MHNC
took account of this by noting that $H_{ei}$ and related  quantities can be
calculated in linear response theory if the interaction potential occurring in $H_{ei}$
could be replaced by a weak pseudopotential $U_{ei}(k)$ but including non-linear short-range
and long-range corrections and also quantum effects via the Kohn-Sham calculation.  Given
the $U_{ei}(k)$, the electron profile $n(k,T_e)$ and the ion profile $\rho(k,T_i)$ caused
by it could be calculated with the linear response functions and hence $<H_{ie}>$ can be evaluated
{\it without} any prescription for a $T_{ie}$ but using a development based on the
Ornstein-Zernike equation~\cite{unpub15}. Subsequently, if needed, the resulting
 $E_{ei}, g_{ei}=n(r)/n_e$ etc., could be examined to obtain a model for $T_{ie}$ if needed. That will
 of course be only a fit parameter without the meaning of a Lagrange multiplier for energy
 conservation, unlike for $T_e, T_i$.

Benedict {\it et al.}~\cite{BeneELR17} have in fact considered how the energy $E_{ei}$ should
 be partitioned between the two subsystems in their recent MD study. Given that their
 $\Gamma_{ii}$ is
typically 12, and $\Gamma_{ee}$ would also range from 12 for $T_e=T_i$ to a factor of 100 smaller,
$\Gamma_{ei}=\sqrt{\Gamma_{ii}\Gamma_{ee}}$ would also range form 0.01 to 12. The $\Gamma_{ie}=12$,
or smaller $\Gamma$  interactions can easily be replaced by a weak NPA-pseuopotential $U_{ei}$ and linear
 response theory may be used to partition $<E_{ei}>$ so that the total energy can be written
 as $E=E_i+E_e$. It is not clear if this will agree with the method used by Benedict 
{\it et al.} where $E_{ei}$ has been 'partitioned equally' between the two subsystems.

In fact, in MD simulations involving $T$-dependent potentials (or otherwise) and with $T/E_f\sim1$,
 the temperature cannot
 be simply estimated form kinetic considerations alone, although such approximations are made in
 simple kinetic models like those of Lenard and Balescu. In the NPA+HNC approach or in any
 similar {\it reduced approach} yielding weak
 pseudopotentials, the pair-potential can be written down correctly in second-order theory. Then the HNC or MHNC equations give the $g_{ii}(r)$. Hence the total free energy $F(T_e,T_i)$
 as well as the component-subsystem
  free energies are explicitly available, even without a coupling constant integration over the
 pair-distribution functions, since one can use Eq. (17) of 
Ref.~\cite{eos95}. This is sufficient for the range of $\Gamma$ used
 in Ref.~\cite{BeneELR17}. The free energies and  specific heats estimated from the MD
 simulations can
be compared with such a  reduced approach and estimates of the temperature that match the
MD can be determined by inverting the data. That is, as long as a reduced approach 
(e.g., NPA + MHNC) can be found, one can give a definite  meaning to the  $T_e,T_i$
 even in a strongly coupled system.

Returning to the question of `dividing' the interaction energy term $F_{12}(T_1,T_2)$ of a
coupled system made up of two subsystems 1 and 2, we can write the Hamiltonian and its free energy
as:
\begin{eqnarray}
H&=&H_1(T_1)+H_2(T_2)+ H_{12}\\
 H_{12}&=&\sum_{k,q} U_{12}(q)n(k)\rho(k+q) \\
F&=&F_1+F_2+F_{12}
\end{eqnarray}
The problem is to partition $F_{12}$ in a meaningful way. We assume that $U_{12}(q)$ has been
constructed to be a weak pseudopotential using a model like the NPA where the non-linear
corrections and effects of the bound-electron core are absorbed into the pseudopotential
which is no longer that of a point ion. In such a case, $F_{12} << F$ is a reasonable assumption,
and we write:
\begin{equation}
f_i=\frac{\exp(-F_i/T_i)F_{12}}{\{\exp(-F_1/T_1)+\exp(-F_2/T_2)\}}
\end{equation}
The individual subsystem free energies $F_i(T_i)$ are known, and this partitions the
interaction free energy $F_{12}$ proportionately. The quantities $f_i(T_i)$ from then
on are assumed to be functions of $T_i$ alone, and the corrected individual subsystem
energies $\tilde{F}_i=F_i+f_i$.  Then the corresponding partitioned contributions
add to the internal energy $E_i$ to give $\tilde{E}_i=-d\{\beta_i(F_i+f_i)\}/d\beta_i$. 
This also enables
a simple estimate of the specific heat and the pressure contributions entirely in terms
of corrections from the  partitioned quantities. The method works as long as $F_{12}$
 is small compared to $F_1,F_2$. Explicit calculations of $F_1, F_2, F_{12}$ for
 2$T$ WDM systems using the
NPA  may be found in Ref.~\cite{Harb17Phon}.

Benedict {\it et al.} give in their Fig.~6 (Ref.~\cite{BeneELR17}) an HNC calculation for
an ion-ion $g(r)$ at $\Gamma_{ii}\sim 12$. One would expect the HNC to agree 
well with the MD, even without any bridge corrections which are quite small here.
 Since a system involving $T_e\ne T_i$,
with $T_e/E_F\simeq 1$  is being used, perhaps this is a system where a properly constructed
classical map for the quantum electrons is needed. The somewhat outdated QSPs used by
Hansen and McDonald
have not been demonstrated to recover  Quantum Monte Carlo (QMC) PDFs or energies for
partially degenerate $T_e/E_F\le 1$ although they perform well at much higher temperatures.
 The use of  coupled HNC equations suggests that Benedict {\it et al.} assume
 a two-thermostat model for their $T_e, T_i$ UFM system.

The classical-map hyper-netted-chain (CHNC) equations~\cite{PDWXC} accurately map
quantum electrons to a {\it classical Coulomb fluid}
 from full degeneracy ($T=0$) to the fully classical
limit, and accurately recover spin-dependent Quantum Monte Carlo (QMC)
 $g_{ee}$ and other quantities at $T=0$. It also recovers the 
path-integral Monte Carlo simulations at finite $T$~\cite{Hungary16,SandipDufty13}
that became available a decade later. 
The attempts to calculate $g_{ee}(r)$ in the degenerate regime using quantum-kinetic
equations had invariably led to $g(r)$ which had unphysical negative regions as soon as the
coupling constant reached even a value of $r_s$=2~\cite{STLS68,IchiIyeTana87}.
 The CHNC was the first model~\cite{CHNC1} that could accurately generate 
 $g_{ee}(r), S_{ee}(k)$ etc., at $T=0$  closely agreeing with QMC, and also yield
$g_{ee}(r)$ and $k, T$-dependent LFCs $G_{ee}(k,T_E)$ at arbitrary spin-polarizations
 and at  finite-$T$. The accuracy of the CHNC $g_{ee}(r,T)$ and other results at finite-$T$ was
 confirmed more than
a decade later by the PIMC calculations of Brown {\it et al}~\cite{BrownXCT13}.

The CHNC can be used for fully ionized hydrogen~\cite{hug02} as well as for more complex
electron-ion systems~\cite{Bredow14}. While the CHNC equations work well for fully ionized systems,
their use with bare Coulomb potentials $Z/r$ where $Z$ is the mean ionic charge is found to
be unsatisfactory for ions with a significant  bound core (e.g., aluminum, Z=3)~\cite{BredowThesis}.  However, the CHNC is easily applicable in the regime free of bound states examined  by studies on energy relaxation. CHNC is as easily implemented as the HNC itself
and accurately includes quantum effects.
\subsection{The case of repulsive ion-electron interactions.}
\label{repuls.sec}
Daligault {\it et al.} have used their equations to interpret molecular-dynamics ER rates
for two subsystems of like charge, i.e., `repulsive hydrogen'. This is a valuable
  idea for obtaining  reliable
simulation results without the need for unphysical cutoffs needed to control attractive 
Coulomb interactions. They have studied `like-charged' classical 'repulsive hydrogen' plasmas 
for $\bar{n}=n_e=1.6 \times 10^{24}$ particles/cm$^3$, i.e, $r_s=r_{ws}$ =1.0,
$\rho=2.63$ g/cm$3$, for $\Gamma=1/(r_sT)$ in the range 0.001 to 1. In a classical
Coulomb plasma calculation only the values of $\Gamma$, $Te, T_i$ are needed in an
 ER rate estimate.

Noting that $V_{ie}=V_{ee}=V_{ii}=Z^2v$, $Z=1$,  $v=V_k$, the {\it cm} denominator
 $\mathcal{D}$ becomes
\begin{eqnarray}
\label{deltaF.eq}
\mathcal{D}&=&1-v\Sigma_j\mathcal{F}_{jj}\chi_{jj}-v\Delta\mathcal{F}\chi_{ee}^0\chi_{ii}^0\\
\Delta\mathcal{F} &=&\mathcal{F}_{ee}\mathcal{F}_{ii}-\mathcal{F}_{ei}\mathcal{F}_{ie}
\end{eqnarray}
Clearly, for repulsive hydrogen, setting $\Delta\mathcal{F}=0$ is a valid approximation
 and we have the simplified form for the denominator:
\begin{equation}
\label{Dsimp.eq}
\mathcal{D}_{sim}= 1-v\Sigma_j\mathcal{F}_{jj}\chi_{jj}
\end{equation}  
The numerator contains the second-order interaction $U_{ei}(k)$, as well as
$A_eA_i\Delta N$, where all the factors are calculated in the independent-subsystem
approximation, as in Eq. (50) of DWP. Any attempt to include terms beyond
the second-order treatment using renormalized quantities (e.g., by replacing 
spectral functions by ones with higher-order corrections) is likely to fail. If such
higher-order terms are
retained, corresponding contributions from the three-vertex diagrams are also needed,
as is well known from theory of the electron liquid~\cite{GT70-pII, RichAsh94}. Hence
we do not attempt to go beyond the 2nd-order form of Eq.~\ref{cm-dwp.eq}.
Our  final form for the ER-rate in `repulsive hydrogen' is given by
\begin{equation}
\label{er-replus.eq}
dE_e/dt=-2\int \frac{dk^3}{(2\pi)}^3 \frac{\omega d\omega}{2\pi}|U_{ei}(k)|^2 
\frac{\Im\chi^0_e\Im\chi^0_i\Delta N_{ei}}{|D_{sim}|^2}
\end{equation}
Thus we see that the RPA approximation of neglecting LFCs is a 
good approximation to the `repulsive-hydrogen' model.
If similar simplifications are carried out on the DD-form of the {\it cm}-ER rate we
obtain an identical equation, except for the differences in specifying
$U_{ei}, U^*_{ei}, G_{jj}, G_{jj}^*, \chi_{jj}^*$ etc., due to our different
 interpretations of the ultra-fast matter system that has to be studied.
The reduced form is convenient for numerical computations,
where $\chi_{ee}$ should be retained as an expansion in $\omega$ near $\omega=0$,
while $\chi_{ee}$ should be approximated by its large-$\omega$ expansion.
However, the pitfalls of such expansions are discussed in the next section.

\subsection{Attempts to simplify the coupled-mode calculation.}
\label{cm-simp.sec}
Several attempts to simplify the coupled-mode calculation, or introduce reduced
alternatives, have appeared in the  literature. Daligault and Mozynsky~\cite{DaliMoz08}
and also Chapman {\it et al.}~\cite{ChapVorb13} have proposed a variant of the {\it cm}-formula
where the ion-ion interaction $V_{ii}(k)$ is screened by the e-e RPA static dielectric
function, leading to terms of the form
\begin{equation}
W=\frac{V_{ii}\chi_{ii}^0}{1-\{V_{ii}/\epsilon_{ee}(k,0)\}\chi_{ii}^0}
\end{equation}
Here $\epsilon_{ee}(k,0)$ is the static electron dielectric function.
It is of course quite impossible to have such screening of the fundamental
interaction $V_{ii}$ as this requires a Dyson equation within a Dyson equation. 
Any insertions of particle-hole loops into an interaction line leads to no
new contributions, and hence it is evident that this is erroneous although
kinetic-equation methods do not have such safe guards as those built into Feynman
techniques.

During our work on the reduction of the ER-rate formula using the $f$-sum
rule as given in Hazak {\it et al.}~\cite{HazakELR01}
the present author tried to
 construct a similar reduced
formula for the cm-ER rate. This was in fact part of our effort
regarding classical constructs for  dealing with quantum electrons interacting
with ions. The first stage of the project was to construct a classical map of
the quantum electrons, which can then be used together with the ions to
generate static quantities like $S_{ee}(k), S_{ei}(k), S_{ii}(k)$. The CHNC
successfully achieves this but does not give dynamic quantities.
But ER-rate calculations need dynamic quantities in addition to static PDFs.
\subsection {Conditions for a conserving approximation}
\label{conserving.sec}
In Ref.~\cite{HazakELR01} a first-order expansion in $\omega$
was used for the relevant response functions. In dealing with {\it cm}-ER one can try such
expansions, retaining the small-$\omega$ regime in $\chi_{ee}$, and the large $\omega$
regime $\chi_{ii}$, and using approximations that preserve the pole structure of the
denominators. However, such expansions in $\omega$ require the satisfaction of
a number of strict conditions.
\begin{enumerate}
\item Since real and imaginary parts are retained after approximation, it is necessary
 to ensure that
the Kramers-Kr\"{o}nig relations are satisfied in some sense.
\item Since we are retaining a finite number of higher-order terms in an $\omega$ expansion, the
frequency-moment sum rules, e.g., to third order, have to be satisfied by suitably readjusting
 the expansion coefficients. A simple example of the need for such adjustment is found already in
the plasmon-pole approximation to the RPA-response function $\chi_{jj}$ which is constructed to
preserve the pole structure $(\omega\pm\omega_k)$ of the inverse dielectric function. But
 it is well known that the form $\Im[\epsilon(k,\omega)]^{-1}\simeq \omega_k[\delta(\omega-\omega_k)
-\delta(\omega+\omega_k)]$ does not satisfy the $f$-sum rule while the modified form
$\Im[\epsilon(k,\omega)]^{-1}\simeq \omega_p[\delta(\omega-\omega_k)
-\delta(\omega+\omega_k)]$ does.

The third-moment sum rule has the form, with $e_q=q^2/2$:\\
\begin{eqnarray}
<\omega^3>&=&\omega_p^2\{e_q^2+4e_q<T/N> +\omega_p^2 J(q)\}, \\
J(q)&=&(1/N)\sum_{k\ne 0}(\frac{(\vec{k}\vec{q})^2}{k^2q^2)}[S(\vec{k}-\vec{q})-S(\vec{k})]
\end{eqnarray}
The third moment involves the mean kinetic energy per particle $<T/N>$ and hence
its satisfaction is needed in a problem involving subsystem temperatures.
Since we need the convenience of treating the LFCs in their static
approximation, the expansion coefficients in powers of $\omega$, chosen to
satisfy the sum rules, will help to overcome the short-comings of usung
static LFCs.

\item The compressibility sum rule has no clear meaning for
non-equilibrium system, but, at least for 2$T$ quasi-equilibrium
systems, one can demand that the subsystem compressibilities,
 calculated from
 the subsystem 2$T$ NPA equation of state agree with the compressibility
obtained from the  $k\to 0$ limit of the subsystem
$S(k)$.
\item The $\chi_{ii}$ obtained from this procedure yields a $S(k)_{ii,m}$ obtained from
 the model. This should agree with the actual $S_{ii}(k)$ obtained from MD, or
form the pair-potential and the MHNC or HNC equation.
\begin{equation}
\label{sk.eq}
S_{ii,m}(k)=\int \{d\omega/2\pi\} \{-2 \Im \chi_{i}(k,\omega)\}. 
\end{equation}
If the model
response function is unsatisfactory, then even the positivity of the $S(k)_{ii,m}$ and the $g(r)$
obtained from it is not guaranteed. In fact, for the $\Gamma_{ii}$ used in Benedict {\it et al.},
the RPA response function (even without any $\omega$-expansion approximations) would fail
 to give a positive
 definite $g(r)$. 
\end{enumerate}

The above scheme was constructed with several of the above conditions imposed via
Lagrange multiplies, and  an attempt was made by the present author to solve for an optimal
set of expansion coefficients in terms of $\omega$ giving a conserving approximation.
 This effort towards the construction of a model {\it cm}-response function was not too
successful and so was not pursued; instead
we concentrated on studying the successful effort with the CHNC calculations for the
static quantities. In fact, the construction of such dynamic approximations for the
response functions have to be undertaken within the context of generating effective potentials,
their static functions like $S(k), g(r)$, and then their phonons, as the
phonon spectrum is closely linked with the ion-ion $S(k,\omega)$, with the longitudinal
branches surviving in the WDM fluid. Our very simple codes achieve this as demonstrated
recently~\cite{Harb17Phon} for room temperature phonons and also for phonons under WDM conditions.
 On the other hand, we do not as yet have a simple dynamical
calculation of $S_{ii}(k,\omega)$ that can be used reliably for ER-rate calculations,
(except for the costly possibility of  MD simulations using the NPA potentials).  

Chapman {\it et al.}~\cite{ChapVorb13} have published a somewhat successful 
reduced {\it cm}-approach where they 
have attempted an $\omega$ expansion and curtailment of the denominators and other relevant
quantities of the response functions entering into the ER-rate calculation.
However, they have not explicitly stated if they treat a system with two thermostats
or not, as this crucially changes the local fields to be used. Furthermore,   the extent of
 validation of the above  sum rules and conditions needed for a conserving
 approximation have not been stated, perhaps because of their greater concern for 
 computational efficiency.
 Instead of checking sum rules, they have opted to check their results by
 directly computing the
full {\it cm}-expression by $\omega, k$ integrations and  claim good agreement in
 some regimes of density and $T_i,T_e$.
In their Fig.1 they have also reported regimes where the {\it cm} contribution is very large.
Surprisingly, this is also the regime where any {\it cm}-modes would be least damped, and hence
not likely to be relevant to energy relaxation. Hence, our suggestion is to check if
the approximations are valid in this regime (e.g., Fig. 1, $n_e> 10^{22}$ cm$^{-3}$
and very high $T_e$) by computing the accuracy of the  previously  mentioned sum rules,
  Kramers-Kr\"{o}nig and other relations. We suspect that those analytical constraints
are probably not well satisfied by their model, and it is likely that the PDFs ($g_{jj'}$))
 calculated from the model response functions of Chapman{\it et al.}
contain regions where they become negative
and hence unphysical.

Benedict {\t et al.}~\cite{BeneELR17} have  also made a frontal
 attack on the problem by doing MD simulations in the regime in question, and
 do not find the large effects
found by Chapman {\it et al.}, although the ER rates obtained from the simulations 
are in fact
 lower than those from the
$f$-sum form of the Fermi Golden rule.

\subsection{Strongly coupled systems}
The method of replacing the electron-ion interaction by a weak pseudopotential
$U_{ei}(k)$ via the Kohn-Sham techniques used in NPA seems to work well even
at very high compressions as far as static properties are concerned, and here
we refer to some recent studies~\cite{Harb17Phon,xrt-LH16, cdw-plasmon16},
 and do not give an exhaustive list of
previous calculations going back to many decades, as they have been summarized
elsewhere~\cite{cdw-cpp15}. Only very few dynamic calculations have been attempted using
NPA potentials~\cite{Nadin88}. 
The phonons calculated from the hottest systems that could be handled by DFT+MD 
agreed quite well with those calculated from the NPA potentials~\cite{Harb17Phon}. This is
a very stringent test of the small-$\omega$, small-$k$ regime of the $S_{ii}(k,\omega)$
that can be obtained from the methods we use. 

In fact, the
regime studied by Benedict {\it et al.} imply $r_{ws}\simeq 0.2525$ a.u., even
though $\Gamma_{ii}$ is moderate. One can envisage a carbon plasma, or carbon
impurities in the H-plasma,  where $Z\sim 6$, and hence
the $\Gamma$ goes to  432. In such systems, the local structure of the ion is
essentially quasi-crystalline, and each ion is `trapped' within a local cage of other ions.
The ions acquire energy by hopping from their cage to another nearby cage where there may
be a lattice-like vacancy. Thus the determinant energy for this process is the {\it Frenkel
frequency} $\omega_{Fr}$ and the corresponding energy $\hbar \omega_{Fr}$. As the ions
 become hot, the hops become more frequent and the particles
 become more moderately coupled, with ions streaming about rather than being locally
 trapped. This can be examined in more detail by a calculation of the Frenkel energy
 as a function of $\Gamma_{ii}$, and expressing ER-rates
via such hopping processes.

The regime of moderate coupling may perhaps be examined  using an approach similar to that of Feynman and Cohen~\cite{FeynCoh56}
where the $S(k,\omega)$ of liquid helium is modeled using the static $S(k)$ and a weakly coupled
excitation spectrum. In our case, given the NPA second-order $U_{ei}(k)$ and its ion-ion
pair potential $U_{ii}(k)$, the $S(k)$ can be obtained accurately using the MHNC equation.
Also, the ion-ion dynamic structure factor under weak coupling but having the
{\it cm}-denominator $\mathcal{D}$ would have ion-excitation poles given by:
\begin{eqnarray}
\omega_i(k)&=&\omega_{pi}/\epsilon_{ee}(k),\; \omega_{pi}=\sqrt(4\pi \rho/M_i)\\
 \epsilon(k)&=&1-V_{ee}(k)(1-G_{ee}(k))\chi_{ee}^0(k)
\end{eqnarray}
This has the behaviour of an acoustic wave for $k<k_c$ and then tends to a relatively
dispersionless value of $\omega_{pi}$ for large $k$. That part of the dispersion is
analogous to the `optical-like' folded branch of the acoustic dispersion of a monoatomic
cubic lattice. Following the spirit
of the  Feynman and Cohen formula,
an approximate form for the ion-acoustic excitation spectrum under strong coupling, and
the dynamic ion-ion structure factor are given by:
\begin{eqnarray}
\varpi_i(k,\omega&)=&\omega_i(k)/S(k), \Gamma_{ik}=\gamma_i(k,\omega)/S(k) \\
\chi_{ii}(k,\omega)&=&Z(k,\omega)/[\{\omega-\varpi_i(k)\}^2+\Gamma_i(k,\omega)] \\ 
S_{ii}(k,\omega)&=&-2 \Im \chi_{ii}(k,\omega)/[1-\exp(-\omega/T_i)]
\end{eqnarray}
The numerator contains an unknown function  $Z(k,\omega)$ whose weak-coupling
form is known.
Adjustable parameters are needed in $Z(k,\omega)$ and in $\varpi_i(k)$ 
to fit the selected form to the sum rules
as discussed earlier, to obtain reliable results. I do not know if some
workers in the WDM community have already tried such an approach or not.

However, the general scheme followed by us proceeds as follows:\\
 (a) We input the
target free-electron density $n_e$, nuclear charge $Z_N$, $T_e$ and  $T_i$
into the NPA-average atom code to output the mean ionization $Z$,
ion density $\rho$, pseudopotential $U_{ei}(k)$, and the pair potential
$U_{ii}(k)$. Phase shifts of continuum states, Kohn-Sham bound states and energies
 are also available at this stage.\\
(A) If $T_e$ is not too high, a DFT+MD calculation using VASP or ABINIT
 is also initiated to compare and confirm the NPA outputs.\\
(b) The NPA potentials are used to generate PDFs and structure factors.\\
(c) They are used to compute basic EOS quantities like the free energies,
    specific heats, compressibilities,  and pressures, both for equilibrium, and
    for 2$T$ quasi-equilibrium systems \\
(d) Dynamical quantities like the phonon spectra, electrical conductivity $\sigma$
     and the X-ray Thomson scattering profiles are calculated, and where possible
    compared with DFT+MD available from (A).\\
(e) If $T_e \ne T_i$, the FGR $f$-sum energy-relaxation rate is calculated 
    assuming a one-thermostat model where $T_i$ is fixed as the low-$T$ subsystem.\\ 
(f) The calculation of the dynamic ion-ion structure factor $S_{ii}(k,\omega)$
    using a generally applicable reduced model has so far not been successful.\\
(g) But good phonon spectra, i.e., $S_{ii}(k,\omega)$ in the harmonic approximation
    for specific $g(r)$ are available.

The steps (a)-(e), (g) are sufficiently simple that they can be done in negligible time
using a small laptop. Steps A and (f) are currently expensive and time-consuming, while
(f) is not in effect available.
 
The possibility of addressing condensed matter physics and
statistical mechanics 
using only pair-distribution functions and density functionals (i.e.,
without  wavefunction calculations) is  discussed in a
more general framework in chapters 8-9 of the book listed in
Ref.~\cite{apvmm13}. 
\section{Conclusion}
We have reviewed the available results on energy relaxation in 2$T$-WDM systems,
starting from our original formulation of the ER- rate problem using the Fermi Golden rule
and the  couple mode forms from two decades ago, and their variants
proposed since then. Numerically the most useful result has been the application of the
$f$-sum to the Fermi Golden rule for the ER rate. The full expressions for the
{\it cm}-form, e.g., those of Daligualt and Dimonte using plasma-kinetic equations, or of
Vorberger and Gerike using the Lenard-Balescu equations, are in  general
agreement with each other for
second-order results using screening interactions, and with those of Dharma-wardana and
Perrot~\cite{DWP-ELR98},  when correctly
interpreted. Sophisticated, demanding  molecular dynamics simulations have been
 carried out  recently, showing that
many brave simplifications of the coupled-mode energy relaxation formula are probably not
reliable for even moderate ion-ion coupling. 
An alternative method besides the MD simulations  for testing proposed
simplifications of the {\it cm}-formula is to determine if the reduced versions satisfy
well-known sum rules adequately.


\begin{thebibliography}{99}
\bibitem{BeneELR17}
Lorin L. Benedict, Michael P. Surh, Liam G. Stanton {\it et al.}, Phys. Rev R {\bf 95}, 043202 (2017).

\bibitem{Ng2011}
Andrew Ng, Int. J. Quant. Chem. {\bf 112}, 150 (2012).


\bibitem{LandauLif10}
E. M. Lifshitz and L. P. Pitaevaskii, {\it Physical Kinetics}, Pergamon, New York (1981).

\bibitem{HanMac83}
J. P. Hansen and J. R. McDonald, Phys. Rev. Lett. {\bf 97A}, 42 (1983).

\bibitem{BoeMoreELR86}
D. B. Boercker and R. M. More, Phys. Rev. A {\bf 33}, 1859 (1986͒).

\bibitem{DWP-ELR98}
M. W. C. Dharma-wardana, and 
F. Perrot, Phys. Rev. E {\bf 58}, 3705 (1998).


\bibitem{Akhiezer75}
A. I. Akhiezer, I. A. Alhiezer, P. V. Polvion, R. V. Polovin, A. G. Sitenko, and K. N. Stephanov,
 {\it Plasma Electrodynamics, Vol. 2,  Nonlinear Theory
 and fluctuations}. Translated by
D. ter Haar.
Pergamon Press, Oxford, England (1975).

\bibitem{NgELR86}   
A. Ng, D. Parfeniuk, P. Celliers, L. Da Silva, R. M. More, and
Y. T. Lee, Phys. Rev. Lett. {\bf 57}, 1595 (1986͒)..

\bibitem{Pozzo11}
Monica Pozzo, 
Michael P. Desjarlais, and Dario Alf\`{e},
Phys. Rev. B, {\bf 84}, 054203 (2011).


\bibitem{pcom}
private communication (2000).

\bibitem{HazakELR01}
G. Hazak, Z. Zinamon, Y. Rosenfeld, and M. W. C. Dharma-wardana
Phys. Rev E {\bf 64}, 066411 (2001).


\bibitem{GerELR05}
D. O. Gericke, J. Phys.: Conf. Series {\bf 11},111 (2005).


\bibitem{VorGerSchELR10}
J. Vorberger, D. O. Gericke, T. Bonath, and M. Schlanges, 
Phys. Rev. E {\bf 81}, 046404 (2010).

\bibitem{DaligDia09}
J\'{e}r\^{o}me Daligault and Guy Dimonte, Phys. Rev. E {\bf 79}, 056403 (2009).

\bibitem{MuriDW-elr08}
M. S. Murillo and M. W. C. Dharma-wardana,
Phys. Rev. Lett. {\bf 100}, 205005 (2008).

\bibitem{Glosli-MMLL08}
J. N. Glosli, F. R. Graziani, R. M. More {\it et al.},
Phys. Rev. E {\bf 78}, 025401(R) (2008).

\bibitem{Gord54}
G. V. Gordeev, Zh. Eksp. Teor. Fiz. {\bf 27}, 19 (1954). 

\bibitem{VorbGeriELR09}
J. Vorberger and D. O. Gerike, Phys. Plasmas, {\bf 16}152002 (2008).

\bibitem{PironBlenski11}
R. Piron and T. Blenski, Phys. Rev. E {\bf 83}, 026403 (2011)
These authors state that  the mean number of free electrons per ion, i.e., $Z$
``..does not correspond to any well-defined observable in the sense of
quantum  mechanics'', i.e, that there is no quantum operator corresponding
to $Z$ and proceed to  reject the use of $Z$. There is also no operator
 corresponding to the temperature or the chemical potential, and many
 other quantities occurring in quantum statistical
mechanics where a heat bath is attached to the system. Thus $T,\mu, Z$ appear as
Lagrange multiplies for the conservation of energy, particle number and charge
conservation in quantum statistical mechanics, and there is no need for
quantum operators for them. Also, more complex Hilbert spaces can be constructed
to accommodate them as operators, e.g., in Thermofield
dynamics introduced by Umezawa. See Ref.~\cite{apvmm13}. 

\bibitem{elrDW01}
M. W. C. Dharma-wardana,  {\it Phys. Rev. E} {\bf 64} 035401 (2001).


\bibitem{KhGly78}
F. C. Khanna and H. R. Glyde, {\it Can. J. Phys.} {\bf 54}, 648 (1978).




\bibitem{PDWXC}
F. Perrot and M. W. C. Dharma-wardana, Phys. Rev. B {\bf 62}, 16536 (2000);
{\it Erratum: } {\bf 67}, 79901 (2003); arXive-1602.04734 (2016).


\bibitem{CDW-elr08}
M. W. C. Dharma-wardana, Phys. Rev. Let. {\bf 101}, 035002 (2008).

\bibitem{gradstein}
I. S. Gradstein, I. M. Ryzhik, {\em Tables of Integrals, Series and products},
 (Academic, New York 1980) Section 8.2, and \S 3351.

\bibitem{br-sin}L. S. Brown, D. L. Preston and R. L. Singleton, Jr., {\it Phys. Rep.} {\bf 410}, 237 (2005).



\bibitem{Harb17Phon}
L. Harbour, M. W. C. Dharmawardana, D. D. Klug, L. J. Lewis,
Phys. Rev. E {\bf 95}, 043201 (2017).
%

\bibitem{Furutani90}
 F. Perrot, Y. Furutani and M.W.C. Dharma-wardana,
Phys. Rev. A 41, 1096-1104 (1990).

\bibitem{Pe-Be}
F. Perrot,  Phys. Rev. E {\bf 47}, 570 (1993).


\bibitem{eos95}
F. Perrot and M.W.C. Dharma-wardana,
Phys. Rev. E. {\bf 52}, 5352 (1995).

\bibitem{Hungary16}
M. W. C. Dharma-wardana,
Current Issues in Finite-$T$ Density-Functional Theory
 and Warm-Correlated Matter. 50$^{th}$ anniversary of Kohn-Sham theory.
Proceedings of the Conference in Density Functional Theory, Debrecen, Hungary, 2016.
Ed. Karlheinz Schwarz and Agnes Nagy. 
Computation  {\bf 4} (2), 16; 2016. 
http://arxiv.org/abs/1602.04734

\bibitem{SandipDufty13}
Dufty, J.; and Dutta, Sandipan; {\it Phys. Rev. E}  {\em 87}, 032101 (2013).

\bibitem{STLS68}
 K.S. Singwi, M.P. Tosi, R.H. Land, and A. Sj\"{o}lander,
Phys. Rev. 176, 589 (1968).

\bibitem{IchiIyeTana87}
S. Ichimaru, H. Iyetomi, and S. Tanaka, Phys. Rep. 149, {\bf 91} (1987).

\bibitem{CHNC1}
M. W. C. Dharma-wardana and F. Perrot, Phys. Rev. Lett. {\bf 84}, 959 (2000).

\bibitem{BrownXCT13}
W. E Brown, J. L. DuBuois, M. Holzmann and D. M. Ceperley,
Phys. Reb. B {\bf 88}, 081102 (2013).

\bibitem{hug02}
M. W. C. Dharma-wardana and  F. Perrot,
 Phys. Rev. B  {\bf 66}, 014110, (2002).


\bibitem{Bredow14}
R. Bredow, Th. Bornath, W.-D. Kraeft, M.W.C. Dharma-wardana and R. Redmer
Contrib. Plasma Phys., No.X, 1–8 (2014). / DOI 10.1002/ctpp.201400080

\bibitem{BredowThesis}
Richard  Bredow, Ph.D Thesis, {\it Berechnung der Struktureigenschaften dichter Plasmen mit der
 Classical-Map Hypernetted Chain Approximation.}, Uinversit\"{a}t Rostock, Germany (2016).

\bibitem{Murillo13}
S. Murillo,
 Jon Weisheit, Stephanie B. Hansen, and M. W. C. Dharma-wardana,
Phys. Rev. E {\bf 87}, 063113 (2013).

\bibitem{Blenski13}
T. Blenski,
 R. Piron, C. Caizergues, B. Cichocki, High Energy Density Physics,
{\bf 9}, 687-695 (2013).


\bibitem{cdw-carbon16}
M. W. C. Dharma-wardana, ArXive [cond-mat] 1607.07511 (2017).

\bibitem{GT70-pII}
D. J. W. Geldart and R. Taylor, Can. J. Phys.{\bf 48}, 167 (1970).
 https://doi.org/10.1139/p70-023,

\bibitem{RichAsh94}
 C. F. Richardson and N. W. Ashcroft, Phys. Rev. B {\bf 50}, 8170 (1994).

\bibitem{DaliMoz08}
J. Daligault and D. Mozynsky,
High Energy Density Physics, {\bf 4}, 58 (2008).


\bibitem{ChapVorb13}
D. A. Chapman, J. Vorberger and D. O. Gerike,
Phys. Rev. E {\bf 88}, 013102 (2013).

\bibitem{xrt-LH16}
L. Harbour, M. W. C. Dharma-wardana, D. Klug and L. Lewis,
Physical Review E {\bf 94}, 053211, (2016).


\bibitem{cdw-plasmon16}
M. W. C. Dharma-wardana,
Phys. Rev. E {\bf 93}, 063205 (2016).
arXiv:1602.04734 (2016).

\bibitem{cdw-cpp15}
M. W. C. Dharma-wardana,
A review of studies on strongly-coupled Coulomb systems since the rise of DFT
and SCCS-1977.
Contrib. Plasma Phys. {\bf 55}, No.2-3, 79-81 (2015).

\bibitem{Nadin88}
 F. Nadin, G. Jacucci and M.W.C. Dharma-wardana,
Phys. Rev. A {\bf 37}, 1025-1028 (1988). 

\bibitem{FeynCoh56}
R. P. Feynman and M. Cohen, Phys. Rev. {\bf 102}, 1189 (1956).

\bibitem{apvmm13}
M. W. C. Dharma-wardana, {\it A physicist's view of
matter and mind}, Ch 8-9, World Scientific, New Jersey (2013).

\bibitem{unpub15}
M. W. C. Dharma-wardana, Unpublished (2015)

\end{thebibliography}
\end{document}